\documentclass[letterpaper, 10 pt, conference]{ieeeconf}  

\IEEEoverridecommandlockouts                              

\usepackage{amsmath,graphicx}
\usepackage{makeidx}
\usepackage{amssymb}
\usepackage{amsmath}
\usepackage{graphicx}
\usepackage{color}
\usepackage[export]{adjustbox}
\usepackage{url}
\usepackage{algorithm}
\usepackage{algorithmic}
\usepackage{graphics}
\usepackage{float}
\usepackage{epstopdf}
\usepackage{subfigure}
\usepackage{epsfig}
\newtheorem{definition}{Definition}
\newtheorem{theorem}{Theorem}

\title{\LARGE \bf
\vspace{-2em}Brain Morphometry Analysis with Surface Foliation Theory$^\S$ 
}

\author{\vspace{-2em}Chengfeng Wen$^{1}$, Na Lei$^{2}$, Ming Ma$^{1,*}$, Xin Qi$^{1}$, Wen Zhang$^{3}$, Yalin Wang$^{3}$ and David Xianfeng Gu$^{1}\vspace{-2em}$  
\thanks{$^\S$This paper is accepted by 2018 EMBC (Engineering in Medicine and Biology Conference).}
\thanks{*Corresponding author}
\thanks{$^{1}$Chengfeng Wen, Ming Ma, Xin Qi and Xianfeng Gu are with Department of Computer Science,Stony Brook University, Stony Brook, NY 11794, U.S.A.
        {\tt\small \{chwen,minma,xinqi,gu\}@cs.stonybrook.edu}}%
\thanks{$^{2}$Na Lei is with School of Software and Technology, Dalian University of Technology, Dalian 116620, China.
        {\tt\small nalei@dlut.edu.cn}}%
\thanks{$^{3}$Wen Zhang and Yalin Wang are with School of Computing, Informatics, and Decision Systems Engineering, Arizona State University, Tempe, AZ 85281, U.S.A.
        {\tt\small \{wzhan139,ylwang\}@asu.edu}}%
}

\begin{document}

\maketitle
\thispagestyle{empty}
\pagestyle{empty}

\begin{abstract}
Brain morphometry study plays a fundamental role in neuroimaging research. In this work, we propose a novel method for brain surface morphometry analysis based on surface foliation theory. Given brain cortical surfaces with automatically extracted landmark curves, we first construct finite foliations on surfaces. A set of admissible curves and a height parameter for each loop are provided by users. The admissible curves cut the surface into a set of pairs of pants. A pants decomposition graph is then constructed. Strebel differential is obtained by computing a unique harmonic map from surface to pants decomposition graph. The critical trajectories of Strebel differential decompose the surface into topological cylinders. After conformally mapping those topological cylinders to standard cylinders, parameters of standard cylinders (height, circumference) are intrinsic geometric features of the original cortical surfaces and thus can be used for morphometry analysis purpose. In this work, we propose a set of novel surface features rooted in surface foliation theory. To the best of our knowledge, this is the first work to make use of surface foliation theory for brain morphometry analysis. The features we computed are intrinsic and informative. The proposed method is rigorous, geometric, and automatic. Experimental results on classifying brain cortical surfaces between patients with Alzheimer's disease and healthy control subjects demonstrate the efficiency and efficacy of our method.

\end{abstract}

\vspace{-2mm}
\section{INTRODUCTION}
\vspace{-2mm}
MRI based brain morphometry analysis has gained extensive interest in the past decades~\cite{winkler2012measuring}. A lot of research works are focused on identifying very early signs of brain functional and structural changes for early identification and prevention of neurodegenerative diseases. Alzheimer's disease (AD), which is the sixth-leading cause of death in the United States, and the fifth-leading cause of death among those age 65 and older as reported by Alzheimer's Association in 2018~\cite{alzheimer20182018}, has received much interest from researchers around the world. Early detection and prevention of AD can significantly impact treatment options, improve quality of life, and save considerable health care costs. As a non-invasive method, brain imaging study has great potential that will powerfully track disease progression and therapeutic efficacy in AD. For example, whole brain morphometry, hippocampal and entorhinal cortex volumes are among most promising candidate biomarkers in structural MRI analysis.  However, missing at this time is a widely available, highly objective brain imaging biomarker capable of identifying abnormal degrees of cerebral atrophy and accelerated rate of atrophy progression in preclinical individuals at high risk for AD in who early intervention is most needed.

Computational geometric methods are widely used in medical imaging fields including virtual colonoscopy and brain morphometry analysis. Rooted in deep geometry analysis research, computational geometric methods may provide rigorous and accurate quantification of abnormal brain development and thus hold a potential to detect preclinical AD in presymptomatic subjects. Specifically, surface morphometry techniques, such as conformal mapping and area preserving mapping, have shown to be feasible and powerful tools in brain morphometry research.

To the best of our knowledge, this is the first work to propose the use of the surface foliation theory for brain morphometry analysis. We validate our method by classifying brain surfaces of patients with Alzheimer disease and healthy control subjects. Experimental results indicate the efficiency and efficacy of our proposed method. The main contributions are summarized as follows:
\vspace{-1mm}
\begin{itemize}
  \item A novel brain surface morphometry analysis method is proposed based on surface foliation theory.
  \vspace{-1mm}
  \item A set of new geometric features (i.e., heights and circumferences) computed by pants decomposition and conformal mapping of topological cylinders are also proposed for surface indexing and classification.
  \vspace{-1mm}
  \item The proposed method is rigorous, geometric and automatic.
\end{itemize}

\vspace{-5mm}
\section{PREVIOUS WORKS}
\vspace{-2mm}
Brain morphometry analysis plays a fundamental role in medical imaging
~\cite{zeng2013teichmuller}. Many research works have investigated the brain morphometry analysis and shape classification.
Winkler et al.~\cite{winkler2012measuring} proposed that the surface area could serve as an important morphometry feature to study brain structural MRI images. Besides, numerous methods have been presented in order to describe shapes, including statistical methods~\cite{osada2002shape}, topology based methods~\cite{hilaga2001topology}, and geometry based methods~\cite{mahmoudi2009three}. To solve real 3D shape problems, researchers have also proposed many shape analysis and classification methods.
Zacharaki et al.~\cite{zacharaki2009classification} proposed the use of pattern classification methods for classifying different types of brain tumors. Recently, Su et al.~\cite{su2015shape} presented a shape classification method using Wasserstein distance. The method computed a unique optimal mass transport map between two measures, and used Wasserstein distance to intrinsically measure the dissimilarities between shapes.

\vspace{-1mm}
Foliation~\cite{strebel1984quadratic} is a generalization of vector field.
In computer graphics field, Zhang et al.~\cite{zhang2006vector} invented a vector field design system which could help users create various vector fields with control over vector field topology. The technique can be used in some applications such as example-based texture synthesis, painterly rendering of images, and pencil sketch illustrations of smooth surfaces. Recently, Campen et al.~\cite{campen2016bijective} proposed a method for bijective parametrization of 2D and 3D objects based on simplicial foliations. The method decomposed a mesh into one-dimensional submanifolds, reducing the mapping problem to parametrization of a lower-dimensional manifold. It was proved that the resulting maps are bijective and continuous. In isogeometric analysis field, Lei et al.~\cite{lei2016quadrilateral} presented a novel quadrilateral and hexahedral mesh generation method using foliation theory. A colorable quad-mesh method was employed to generate the quadrilateral mesh based on Strebel differentials, which then leads to the structured hexahedral mesh of the enclosed volume for high genus surfaces.

\section{THEORETIC FOUNDATION}
\vspace{-2mm}
\label{sec:theory}
We briefly review the basic concepts and theorems in conformal geometry. Detailed treatments can be found in~\cite{gu:08:ComputationalConformalGeometry}.

If a complex function $f:\mathbb{C}\to \mathbb{C}$ $(x,y)\to(u,v)$ satisfies the Cauchy-Riemann equation
\vspace{-3mm}
\[
u_x = v_y, u_y = -v_x,\vspace{-2mm}
\]

then $f$ is called a \emph{holomorphic function}. If $f$ is invertible, and $f^{-1}$ is also holomorphic, then $f$ is a \emph{bi-holomorphic function}. A surface with a complex atlas $\mathcal{A}$, such that all chart transition functions are bi-holomorphic, then it is called a \emph{Riemann surface}, and the atlas $\mathcal{A}$ is called a \emph{complex structure}. All oriented metric surfaces are Riemann surfaces.

\vspace{-1mm}
\begin{definition}[Holomorphic Quadratic Differentials\newline]
Let $S$ be a Riemann surface, and $\Phi$ a complex differential form, such that on each local chart with the local complex parameter $\{z_\alpha\}$,
$
    \Phi = \varphi_\alpha(z_\alpha) dz_\alpha^2,
$
where $\varphi_\alpha(z_\alpha)$ is a holomorphic function. The $\Phi$ is called a holomorphic quadratic differential.
\end{definition}
\vspace{-1mm}

According to Riemann-Roch Theorem, the linear space of all holomorphic quadratic differentials is $3g-3$ complex dimensional, where the genus $g>1$.
A point $z_i \in S$ is called a \emph{zero} of $\Phi$, if $\varphi(z_i)$ vanishes. A holomorphic quadratic differential has $4g-4$ zeros. For any point away from zero, we can define a local coordinates
\vspace{-2mm}
\begin{equation}
    \zeta(p) := \int^p \sqrt{\varphi(z)}dz.
    \label{eqn:natural_coordinates}
    \vspace{-2mm}
\end{equation}
which is the so-called \emph{natural coordinates} induced by $\Phi$. The curves with constant real (imaginary) natural coordinates are called the \emph{vertical (horizontal) trajectories}. The trajectories through the zeros are called the \emph{critical trajectories}.

\vspace{-1mm}
\begin{definition}[Strebel]Given a holomorphic quadratic differential $\Phi$ on a Riemann surface $S$, if all of its horizontal trajectories are finite, then $\Phi$ is called a Strebel differential.
\label{def:strebel_differential}
\end{definition}
\vspace{-1mm}

A holomorphic quadratic differential $\Phi$ is Strebel, if and only if its critical horizontal trajectories form a finite graph \cite{strebel1984quadratic}.
The Strebel differentials are dense in the space of all holomorphic quadratic differentials. Given a holomorphic quadratic differential $\Phi$, the natural coordinates in Eqn.~\ref{eqn:natural_coordinates} induce a flat metric with cone singularities (cone angles equal to $-\pi$), which is denoted as $|\Phi|$.  Hubbard and Masur proved the following existence of a Strebel differential with prescribed type and heights.

\vspace{-1mm}
\begin{theorem}[Hubbard and Masur]Given non intersecting simple loops $\Gamma=\{\gamma_1,\gamma_2,\cdots, \gamma_{n}\}$, and positive numbers $\{h_1,h_2,\cdots,h_{n}\}$, $n\le 3g-3$, there exists a unique holomorphic quadratic differential $\Phi$ satisfying:
\vspace{-1mm}
\begin{enumerate}
\item The critical graph of $\Phi$ partitions the surface into $n$ cylinders, $\{C_1, C_2,\cdots, C_{n}\}$, such that $\gamma_k$ is the generator of $C_k$,
\item The height of each cylinder $(C_k,|\Phi|)$ equals to $h_k$, $k=1,2,\cdots,n$.
\end{enumerate}
\label{thm:existence_strebel}
\end{theorem}
\vspace{-1mm}

The geometric interpretation of above theorem is as follows: each cylinder $C_k$ becomes a canonical flat cylinder under $|\Phi|$, whose height is $h_k$. Therefore, Strebel's theorem allows one to specify the type of $\Phi$ and the height of each cylinder $C_k$.\\

\vspace{-3mm}
\noindent\textbf{Harmonic Map}
Suppose $G=\langle E,N\rangle$ is a graph, $\mathbf{h}:E\to \mathbb{R}^{+}$ is an \emph{edge weight} function. Let $p$ and $q$ be two points on the graph, and $d_{\mathbf{h}}(p,q)$ be the shortest distance between them. Let $(S,\mathbf{g})$ be a surface with a Riemannian metric $\mathbf{g}$. Consider a map $f:(S,\mathbf{g})\to (G,\mathbf{h})$. We say a point $p\in S$ is a \emph{regular point}, if its image is not any node of $G$, otherwise a \emph{critical point}. The set of all critical points is denoted as $\Gamma$. For each regular point $p\in S$, we can find a neighborhood $U_p$, and the restriction of the map on $U_p$ can be treated as a normal function $f:U_p\to \mathbb{R}$. We choose an \emph{isothermal coordinates} $(x,y)$ on $U_p$, such that the metric has a special form $\mathbf{g}=e^{2\lambda(x,y)}(dx^2+dy^2)$. The harmonic energy is given by
$
    E(f|_{U_p}):= \int_{U_p} |\nabla_{\mathbf{g}} f|^2 dA_\mathbf{g}
    \vspace{-1mm}
$,
where $ \nabla_{\mathbf{g}} = e^{-\lambda} \left(\frac{\partial }{\partial x}, \frac{\partial }{\partial y}\right)^T$, and the area element is $dA_\mathbf{g} = e^{2\lambda} dxdy$. The \emph{harmonic energy} of the whole map is defined as
\vspace{-2mm}
\[
    E(f) := \int_{S\setminus \Gamma}  |\nabla_{\mathbf{g}} f|^2 dA_\mathbf{g}.\vspace{-2mm}
\]
The critical point of the harmonic energy is called a \emph{harmonic map}. Wolf \cite{Wolf96onrealizing} proved the existence and the uniqueness of the harmonic map.
\vspace{-1mm}
\begin{theorem}[Wolf] In each homotopy class, the harmonic map $f:(S,\mathbf{g})\to (G,\mathbf{h})$ exists and is unique. Furthermore, the Hopf differential $\Phi=\langle f_z,f_z\rangle dz^2$ induced by the harmonic map is a holomorphic quadratic differential, where $z=x+iy$ is the complex isothermal coordinates of $(S,\mathbf{g})$.
\label{thm:wolf}
\end{theorem}

\noindent\textbf{Conformal Module} Suppose $(S,\mathbf{g})$ is a surface of genus $g>1$. Given $3g-3$ non-intersecting simple loops $\Gamma=\{\gamma_i\}$ and positive numbers $\{h_i\}$, the unique Strebel differential $\Phi$ in Hubbard and Masur's theorem induces a flat metric $|\Phi|$ with cone singularities, and cylinders $\{C_i\}_{i=1}^{3g-3}$. The height and circumference of each cylinder $(C_k,|\Phi|)$ are ($h_k$, $l_k$). The set of all ($h_k$, $l_k$) are the conformal modules.

\vspace{-1mm}
\section{ALGORITHM}

\begin{figure}[!t]
\centering
\includegraphics[width=0.5\textwidth]{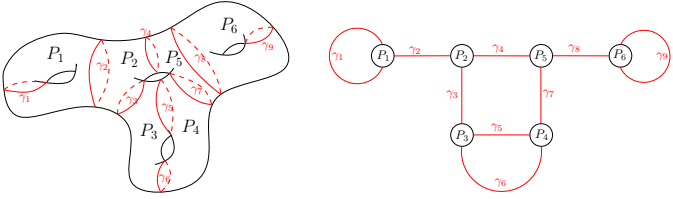}
\caption{Pants decomposition of surface (left) and pants decomposition graph (right).\vspace{-4mm}}
\vspace{-3mm}
\label{fig:pants}

\end{figure}

\vspace{-2mm}
\noindent\textbf{Pants Decomposition}\label{pants-decomposition}
Let $S$ be a closed surface of genus $g$, represented by triangular mesh. Let $\Gamma = \{\gamma_i, i=1,2,...,3g-3\}$ be a set of \textit{admissible curves},  which can be generated automatically or manually specified. User also specifies a height parameter $h_i$  for each admissible curve $\gamma_i$. These admissible curves decompose surface $S$ to a set of \textit{pants} $\mathcal{P} = \{P_i, i=1,2,...,2g-2\}$. The pants decomposition graph $G$ is then constructed in the following way:
\begin{itemize}
\item each pants $P_i$ corresponds to a node in $G$
\item each admissible curve connecting two pants corresponds to an edge  in $G$; two pants may be the same, in that case, the edge becomes a loop
\end{itemize}
Fig.~\ref{fig:pants} illustrates pants decomposition and pants decomposition graph.

\vspace{-2mm}

\begin{figure}[!t]
  \centering
  \begin{tabular}{ccc}
  \includegraphics[valign=m,width=0.14\textwidth]{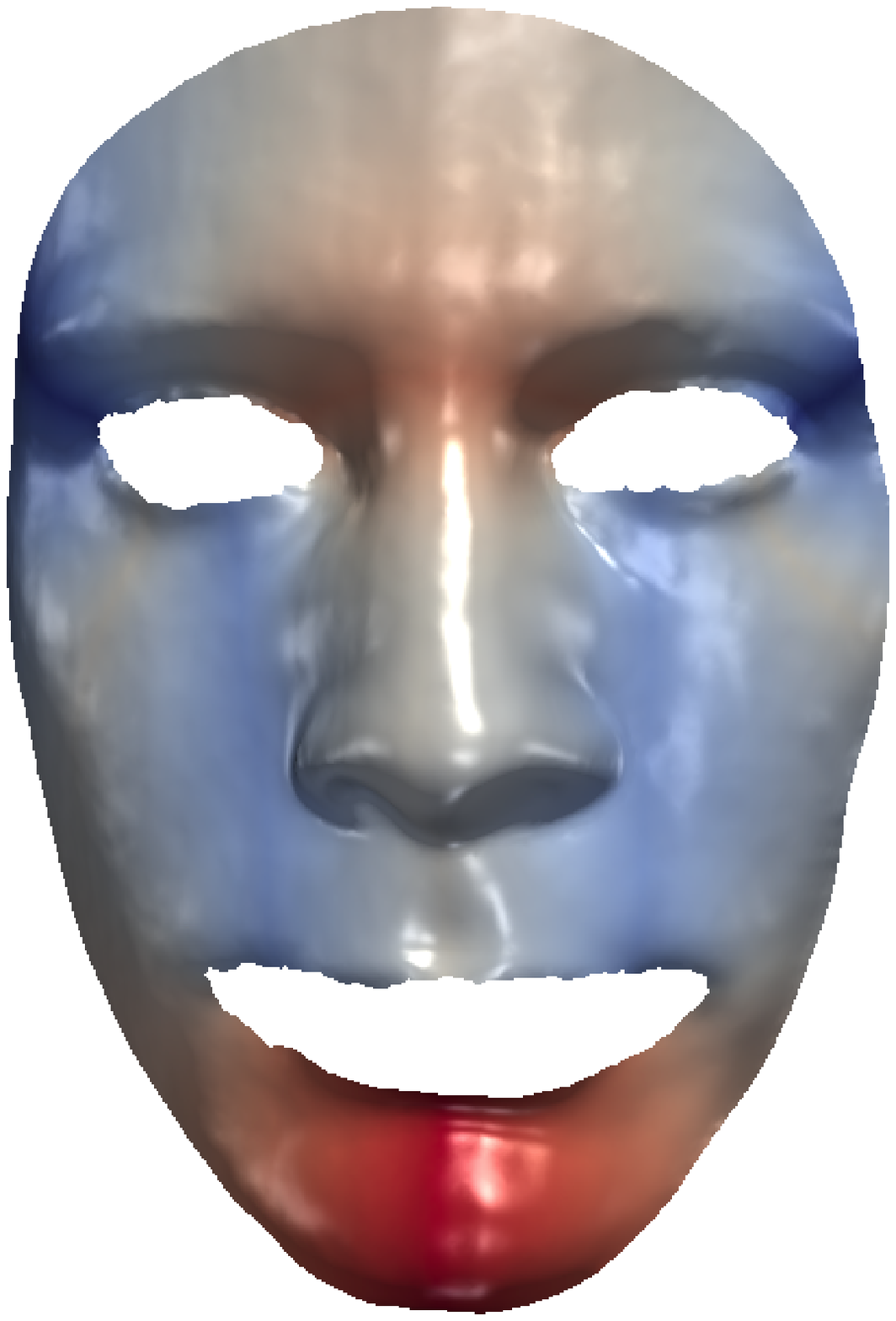}&
  \includegraphics[valign=m,width=0.14\textwidth]{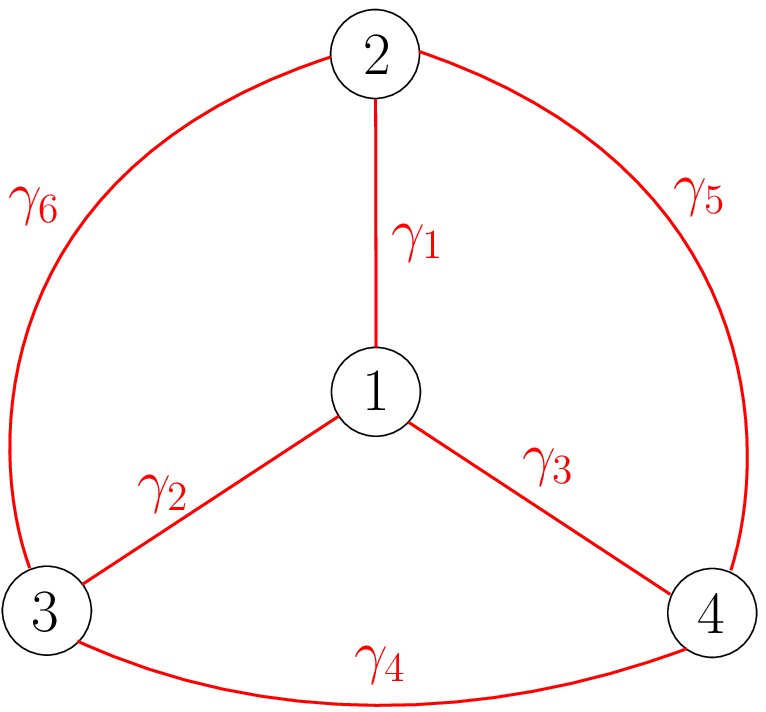}&
  \includegraphics[valign=m,width=0.14\textwidth]{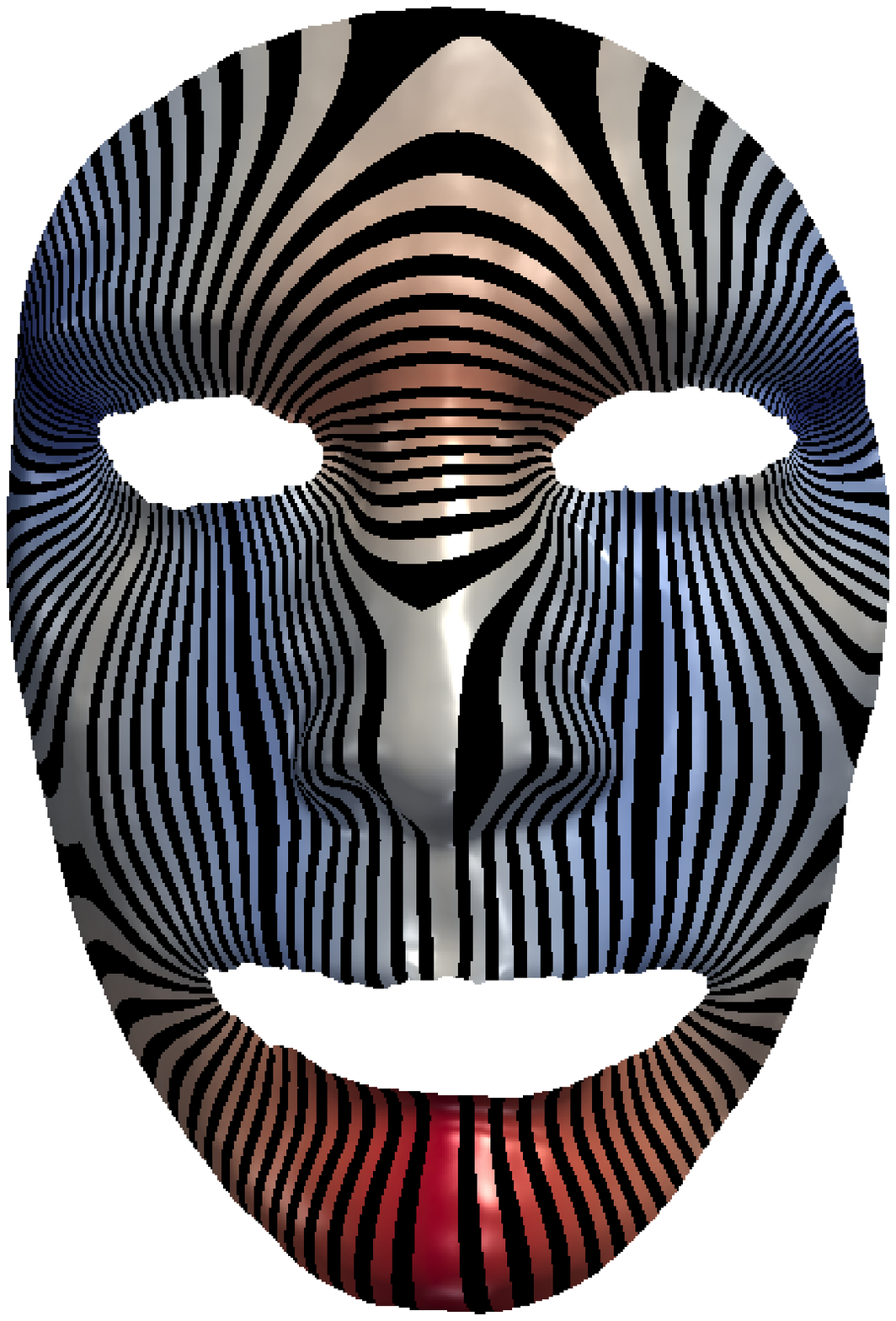}\\
  (a)  & (b) & (c) \vspace{-2mm}
  \end{tabular}
  \caption{Harmonic map from human face to pants decomposition graph and induced surface foliation.\vspace{-8mm}}
  \vspace{-4mm}
  \label{fig:harmonic}
\end{figure}

\noindent\textbf{Discrete Harmonic Map to Graph}
We compute a unique harmonic map $f$ from surface $S$ to $G$. The harmonic energy is defined as
\vspace{-4mm}
\[
E(f) = \sum_{i,j,e_{ij} \in S}w_{ij}d^2(f(v_i),f(v_j))\vspace{-3mm}
\]
where $v_i$ is vertex on $S$, $f(v_i)$ on $G$, $e_{ij}$ is edge, and $w_{ij}$ is cotangent weight.

For each $v_i$, by moving $f(v_i)$ to the barycenter of its neighbors on graph $G$, the energy $E$ will decrease monotonically, which is due to the following definition of barycenter. By iteratively doing so, the energy $E$ will attain its minimum value, at which point we obtain a harmonic map  $f: S \to G$. Thm.~\ref{thm:wolf} guarantees this harmonic map we obtained is the unique one. Fig.~\ref{fig:harmonic} (a) and (b) illustrate harmonic map from a human face surface (a) to its pants decomposition graph (b), and Fig.~\ref{fig:harmonic} (c) shows the surface foliation, where color indicates vertices' target position on graph $G$.

The initial map $f_0$ should be specified in the same homotopy class as the final harmonic map $f$. Subgraph at a node consists of the node and all edges connecting to it. Then initial map can be obtained automatically in the following way: each pants $P_i$ is mapped to the subgraph $G_i$ at node $i$ of $G$, then all pants maps are glued together to obtain $f_0$.

\noindent\textbf{Barycenter Calculation}
For each $f(v_i)$, we move $f(v_i)$ to the barycenter of its neighbors. Calculating barycenter is done by minimizing energy
\vspace{-2mm}
\[
f(v_i)^* = \arg\min_{f(v)} \sum_{j, e_{ij} \in S}w_{ij}d^2(f(v),f(v_j))\vspace{-2mm}
\]
where the right are exactly the terms in $E$ that involve $f(v_i)$. $d(f(v_i),f(v_j))$ can be calculated piecewisely. Then minimization of above energy boils down to minimum calculation of a set of quadratic functions.

\noindent\textbf{Surfaces with Boundaries}
For surfaces with boundaries, we can either double cover those surfaces to obtain a closed surface, or we can add boundaries to the set of admissible curves, and such curves correspond to open edges on $G$. Computation of harmonic map remains the same.

\smallskip

\noindent\textbf{Extract Geometric Features}
A holomorphic quadratic differential $\Phi$ can be induced from the harmonic map we obtained. Tracing the critical trajectories of $\Phi$ and slicing surface along them, we obtain a set of $3g-3$ topological cylinders, each corresponding to an input admissible curve. The set of heights and circumferences of those cylinders are topological invariants, which we propsose to use as geometric features for classification problems in next section.

\vspace{-2mm}
\section{EXPERIMENT}

\begin{figure*}[!t]
\centering
\vspace{-2mm}
\includegraphics[width=1\textwidth]{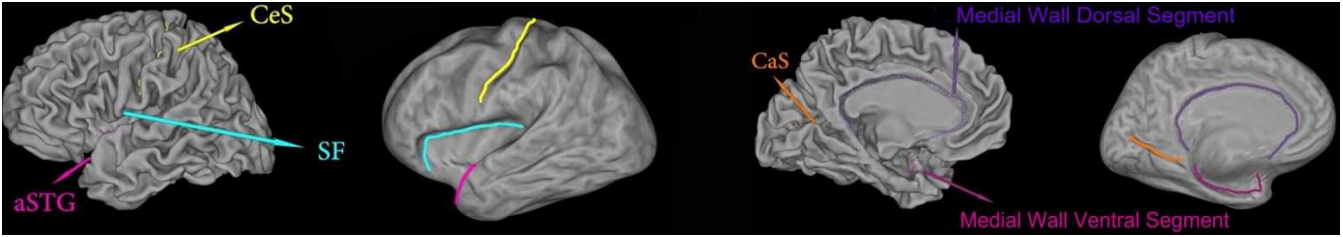}
\vspace{-6mm}
\caption{A left cortical surface with six landmark curves, which are automatically labeled with Caret, showing in two different views on both the original and inflated surfaces.}
\vspace{-6mm}
\label{fig:landmarks}
\end{figure*}

\begin{figure}[!t]
\begin{center}
\begin{tabular}{cc}
\epsfig{file=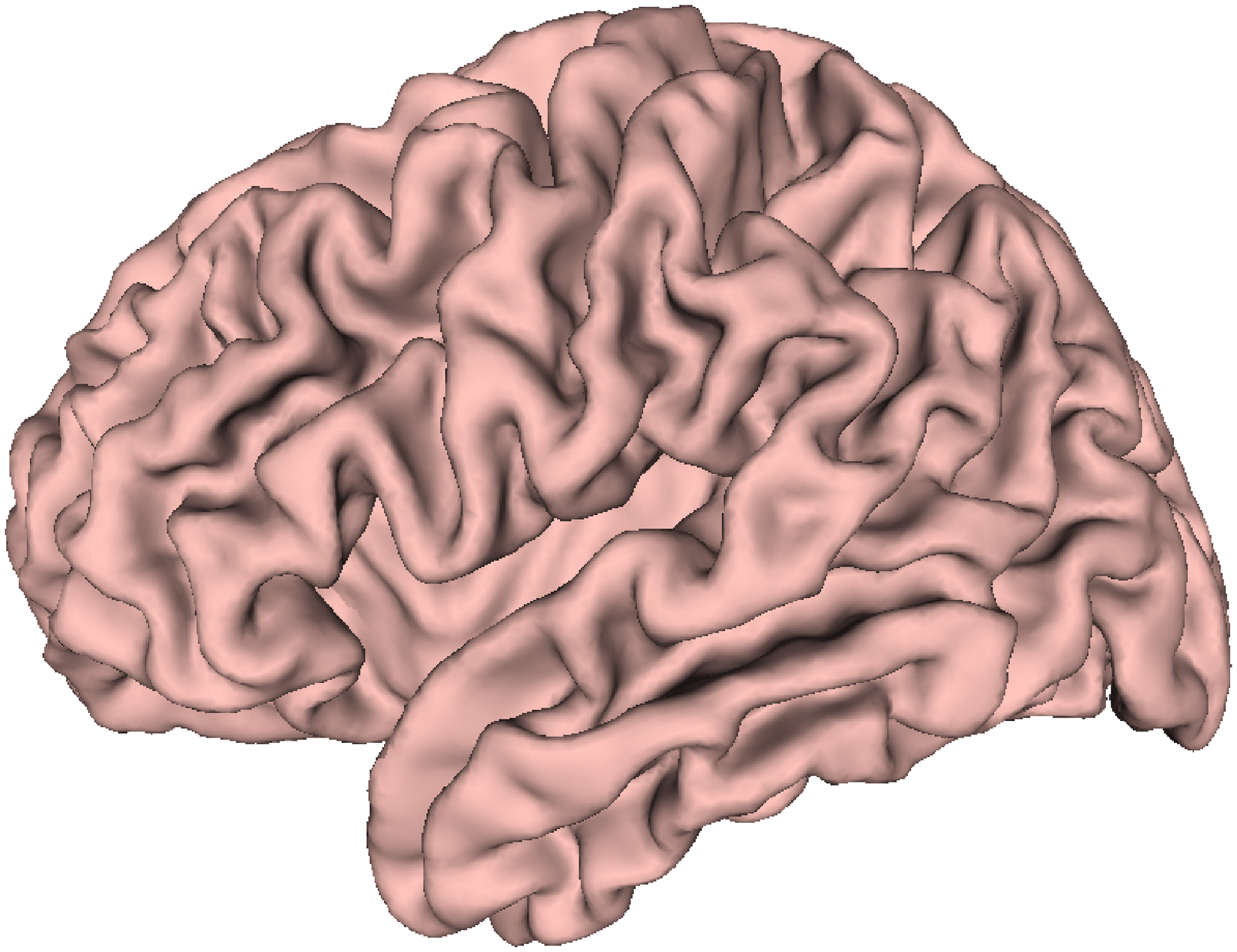,width=0.25\textwidth}&
\epsfig{file=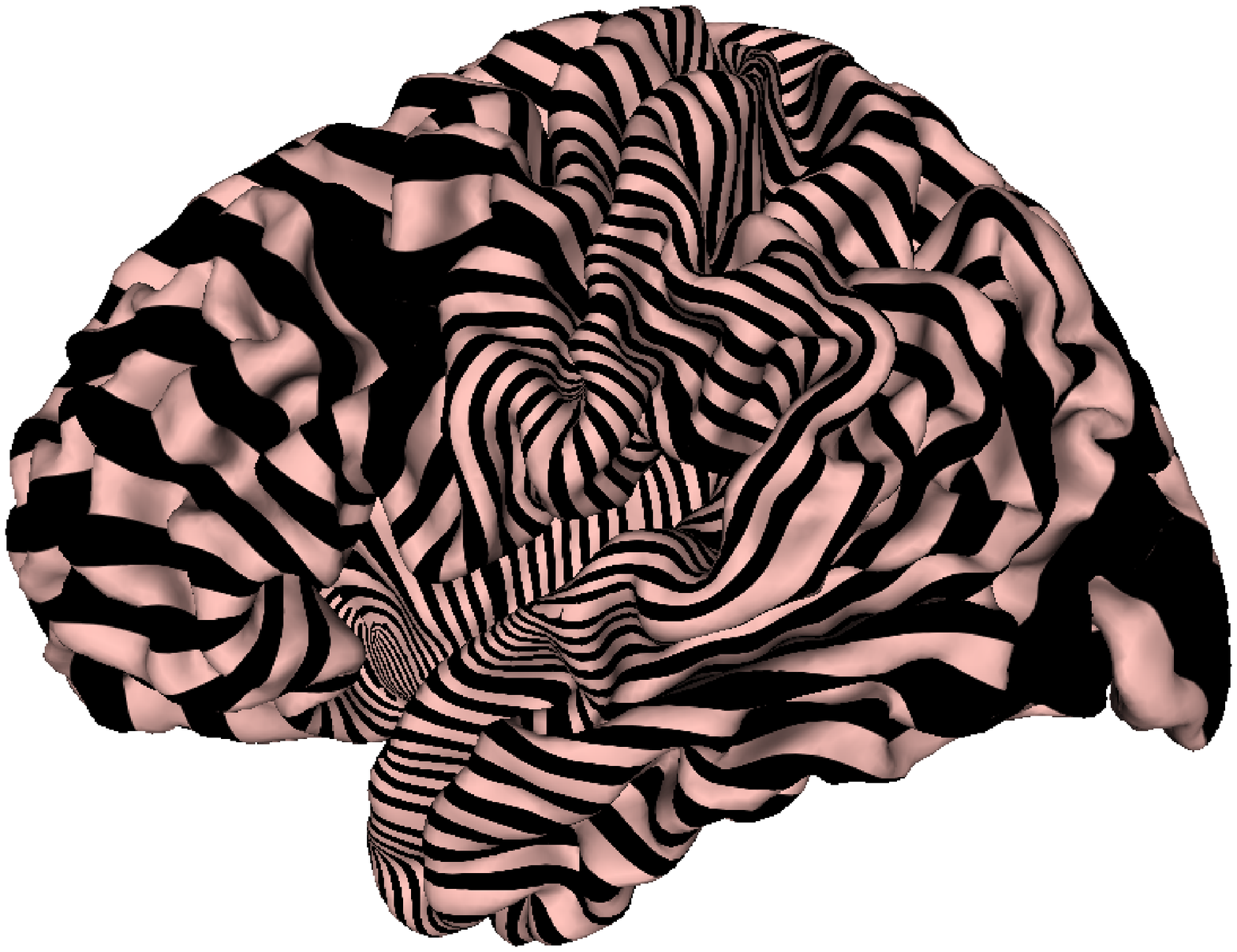,width=0.25\textwidth}\\
(a)  & (b)\vspace{-2mm}
\end{tabular}
\end{center}
\vspace{-2mm}
\caption{A brain surface and its foliation.\vspace{-6mm}}
\label{fig:texture}
\vspace{-2mm}
\end{figure}

\noindent\textbf{Data Preparation}
The dataset used in our experiments includes images from 30 patients with Alzheimer disease (AD) and 30 healthy control subjects. The structural MRI images were from the Alzheimer's Disease Neuroimaging Initiative (ADNI)~\cite{mueller2005alzheimer}. The brain cortical surfaces were reconstructed from the MRI images by FreeSurfer.  Then, a set of `Core 6' landmark curves, including the Central Sulcus (CeS), Anterior Half of the Superior Temporal Gyrus (aSTG), Sylvian Fissure (SF), Calcarine Sulcus (CaS), Medial Wall Ventral Segment, and Medial Wall Dorsal Segment, are automatically traced on each cortical surface using the Caret package~\cite{caret}. In Caret software, the PALS-B12 atlas is used to delineate the ``core 6'' landmarks, which are well-defined and geographically consistent, when compared with other gyral and sulcal features on human cortex. The stability and consistency of the six landmarks were validated in \cite{VanEssen:NIMG05}. An illustration of the landmark curves on a left cortical surface is shown in Fig.~\ref{fig:landmarks} with two views. We show the landmarks with both the original and inflated cortical surfaces for clarity. A brain surface and its foliation are shown in Fig.~\ref{fig:texture} (a) and (b), respectively.

The experimental procedures involving human subjects described in this paper were approved by the Institutional Review Board.

\begin{figure}[!t]
\centering
\includegraphics[width=0.25\textwidth]{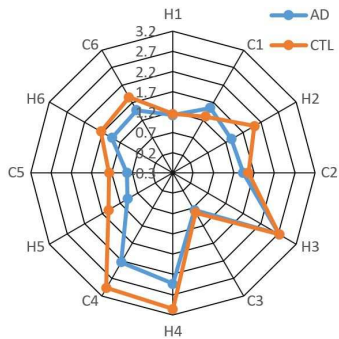}
\vspace{-2mm}
\caption{Radar chart.\vspace{-6mm}}
\vspace{-2mm}
\label{fig:radar_chart}
\end{figure}

\noindent\textbf{Foliation Feature Visualization} We illustrate the difference of feature values between a pair of subjects with AD and healthy control subject (CTL) using radar chart. Radar chart displays multi-variate data in a two-dimensional chart where multiple variables are represented on axes starting from the same point. As shown in Fig.~\ref{fig:radar_chart}, six pairs of heights(H) and circumferences(C) corresponding to "core 6" landmarks, i.e., twelve features (labeled by 'H1', 'C1',...,'H6', 'C6') are associated with twelve corners on the radar chart. We find that the pair of the H4 height and C4 circumference features associated to landmark curve of medial wall dorsal segment have the largest difference between these two subjects' radar charts represented by a blue color line and an orange color line respectively. Although more validations are warranted, our research results may help discover AD related brain atrophy patterns.

\noindent\textbf{Classification}
We validated our method with brain surface classification on a dataset of brain cortical surfaces from 30 patients with Alzheimer disease and 30 healthy control subjects. The SVM method was employed as the classifier with 10-fold cross validation in our experiments. For each image, the input feature vector of the classifier includes 12 features. For comparison purpose, we also computed cortical surface area and cortical surface mean curvatures, two cortical surface features frequently adopted in prior structural MRI analyses~\cite{pmid25100588}. We also applied SVM as the classifiers for these two features. Experimental results are shown in Table~\ref{tbl:comparison_experiment}. Our proposed method achieved 78.33\% correctness rate, which is better than the correctness rate 56.67\% in the brain surface area based method and 55.00\% in the brain surface mean curvature based method. Although multi-subject studies are clearly necessary, this experiment demonstrates that the foliation theory based geometric features may have the potential to quantify and measure AD related cortical surface changes.

\vspace{-2mm}
\setlength{\tabcolsep}{1mm}

\setlength{\tabcolsep}{1mm}

\begin{table}[!t]
\centering
\caption{Classification accuracy comparison between our method and other methods.\vspace{-2mm}}
\label{tbl:comparison_experiment}
\begin{tabular}[\textwidth]{@{\extracolsep{\fill}}|l|c|}
\hline
Classification method &  Correctness rate \\
\hline
Our Method  & \textbf{78.33}\%  \\
Brain Surface Area &  56.67\%  \\
Brain Mean Curvature & 55.00\% \\
\hline
\end{tabular}
\vspace{-6mm}
\end{table}

\section{CONCLUSION}
\vspace{-2mm}
In this paper, a novel set of brain surface features are proposed based on surface foliation theory. To validate our method, we applied our method on classifying brain cortical surfaces between AD and healthy control subjects, and the preliminary experimental results demonstrated the efficiency and efficacy of our method. In future, we will employ our method to explore brain morphometry related to mild cognitive impairment (MCI) and other medical imaging applications.\vspace{-4mm}

\bibliographystyle{IEEEtran}
\bibliography{references}
\vspace{-2mm}

\end{document}